\documentclass[lettersize,journal]{IEEEtran}
\usepackage{amsmath,amsfonts}
\usepackage{algorithmic}
\usepackage{algorithm}
\usepackage{array}
\usepackage[caption=false,font=normalsize,labelfont=sf,textfont=sf]{subfig}
\usepackage{textcomp}
\usepackage{stfloats}
\usepackage{url}
\usepackage{verbatim}
\usepackage{graphicx}
\usepackage{cite}
\usepackage{amssymb}
\usepackage{mathtools}
\usepackage{arydshln}

\usepackage{xcolor}

\begin{document}
\title{Bertrand's Representation of the Optimal Detector}
\author{Vladimir Lenok 
\thanks{The author is with
Fakult\"{a}t f\"{u}r Physik, Universit\"{a}t Bielefeld, Postfach 100131, 33501 Bielefeld, Germany
(e-mail: vlenok@physik.uni-bielefeld.de)}}

\markboth{IEEE Transactions on Signal Processing,~Vol.~XXX, No.~XXX, August~XXXX}
{Lenok: Bertrand's Representation of the Optimal Detector}

\IEEEpubid{0000--0000/00\$00.00~\copyright~2021 IEEE}

\maketitle

\begin{abstract}
It is shown how the optimal detector of Gaussian signals can be represented in terms of Bertrand's class of time-frequency distributions.
In this representation, the detector is a correlation between the corresponding time-frequency distributions.
Since Bertrand's class is related to the power-law chirp signals, the new representation can be useful for their detection.
The new approach is shown to be more effective then other time-frequency methods for the case of phase-insensitive detection.
The finding provides a complementary representation to Cohen's class representation in the time-frequency domain already known in the literature.
\end{abstract}


\section{Introduction}
\IEEEPARstart{O}{ptimal}
detection of a signal in the Bayesian sense is a classically well-studied topic in the field of signal processing with no shortage of historically robust results.
In this work, we add to this list by examining the topic in the context of the power-law chirp signals detection.

\begin{figure}[!t]
\centering
\includegraphics[width=1.\linewidth]{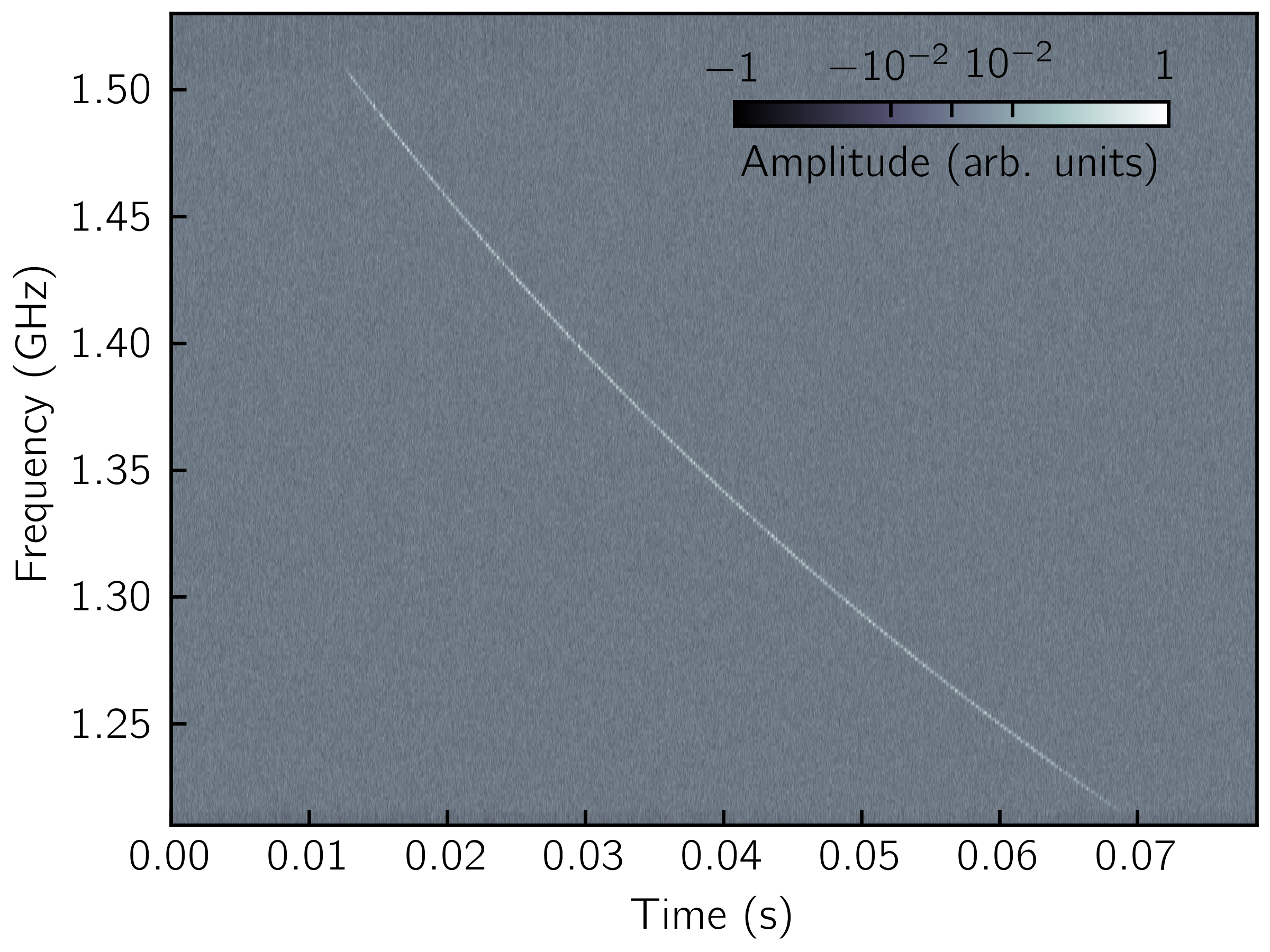}
\caption{
A spectrogram of a bright pulse of the Crab Pulsar detected by the Effelsberg 100-m Radio Telescope.
The apparent dispersion of the pulse is due to the propagation through the interstellar medium.
For presentation purposes the signal is shown in the symmetric logarithmic scale (the region from $-10^{-2}$ to $10^{-2}$ is linear).
Image provided by Andrei Kazantsev, MPIfR.}
\label{fig-pulse}
\end{figure}

In radio astronomy, power law chirps are often of interest in the detection of weak signals from pulsars, fast rotating neutron stars emitting beams of stochastic radio emission towards Earth~\cite{lorimer-kramer-handbook, condon-ransom-book}.
As these signals propagate through the plasma of the interstellar medium, they are dispersed in such a way that they have a group delay
\begin{equation}
    t_X(f) = \frac{\mathcal{D}}{f^2}
    \label{eq:group-delay}
\end{equation}
where the constant $\mathcal{D}$ is proportional to the amount of electrons in the plasma along the line of propagation~\cite{lorimer-kramer-handbook, condon-ransom-book}.
Fig.~\ref{fig-pulse} shows a typical example of such a signal.
Taking into account the stochastic nature of these signals and the functional form of the group delay, it seems reasonable to classify them formally as the Gaussian power-law chirps.

The contemporary method used to detect such signals is the spectrogram correlation.
From a technical perspective, a filter bank splits the original signal into multiple spectral channels, forming a sonograph, which in turn are averaged over small time intervals to reduce the amount of data.
The resulting averaged sonograph is used for further analysis~\cite{2022MNRAS.513.1386V, epta-data, nanograv-pulsar-timing, nanograv-single-pulse}.
Despite their perspective technical implementation sonographs and spectrograms are equivalent in most practical cases (see theorem 2.4.1 in~\cite{boashash2016} showing the equivalence of spectrograms and sonographs and~\cite{meerkat-correlator, lofar-correlator-proceeding, lofar-filterbank} for particular implementations of their use in modern radio astronomical instrumentation).

Although detection of detection of pulsar signals is a routine procedure in radio astronomy that has brought invaluable scientific results, it is known that the method of spectrogram correlation is not optimal~\cite{flandrin1988}.

In this work, we approach the problem of Gaussian power-law chirp detection by finding an alternative to the spectrogram correlation method in the time-frequency domain that can form an optimal detector.
The time-frequency formulation of the optimal detection was already considered in~\cite{kay_optimality_1985, flandrin_detection-estimation_1986, flandrin_time-frequency_1988, papandreou_use_1994, papandreou_detection_1994}.
The most holistic approach is presented in~\cite{flandrin1988}.
It enables the consideration of a wide range of time, frequency, and time-frequency detectors within the same theoretical framework, and it highlights the requirements time-frequency distributions must satisfy to serve as optimal detectors.
The central idea of the time-frequency optimal detector in~\cite{flandrin1988} is to find a representation of the optimal detector in terms of Cohen's class of time-frequency distributions.
Despite the seemingly general character of the solution it seems to be inherently related to linear chirps, since the Wigner-Ville distribution, the main component of Cohen's class, perfectly localizes them on the time-frequency plane~\cite{cohen-review}.
\IEEEpubidadjcol

In contrast to Cohen's class, Bertrand's class of time-frequency distributions is seemingly related to power-law chirps, since it is capable to perfectly localize these type of signals ~\cite{bertrand1992}.
Thus, it seems reasonable to find an optimal detector based on that class.
A time-frequency detector based on Bertrand's class was already presented in~\cite{chassande-mottin1999}.
However, this version is not optimal for a general case of Gaussian power-law chirps, since it is centered around a concept of path integration in the time-frequency plane that may not be suitable for all cases.

The purpose of this paper is to extend the ideas expressed in~\cite{flandrin1988, chassande-mottin1999} and to find the time-frequency optimal detector for Gaussian power-law chirps by finding the representation of the optimal detector in terms of Bertrand's class.

The paper is organized in the following way.
Section \ref{sec:bert} recalls the definition of Bertrand's class and its properties relevant for the detection problem.
By definition, Bertrand's class uses analytic signals, not just complex signals.
It is indeed this fact which complicates the task of finding Bertrand's representation of optimal detector for Gaussian signals.
Therefore, we must first develop the optimal detector for analytic signals from first principles.
In this way, in Section \ref{sec:formulation}, we must first clearly specify the detection problem.
This problem is then solved throughout the sections \ref{sec:opt-det-general}--\ref{sec:opt-det-analytical}.
In section \ref{sec:opt-det-bert}, the resulting optimal detector for the analytic signals is represented in terms of Bertrand's class.
Section \ref{sec:examples} provides illustrative examples of the specific variants of the new representation.

\section{Bertrand's class of time-frequency distributions}
\label{sec:bert}
Bertrand's class of time-frequency distributions was originally developed for wide-band radar applications and
has been extensively studied over the last decades~\cite{bertrand1992, boashash2016}.
Here, we briefly review some aspects of this class relevant for the further development of representation of the optimal detector.

For an analytic signal $X(f)$, given in the frequency domain, Bertrand's class of time-frequency distributions is defined as the integral transform
\begin{align}
\label{eq:bertrand}
  B(t, f) =
  f^p
  &\int_{-\infty}^{\infty}
  X\bigl(f\lambda(u)\bigr)\,
  X^*\bigl(f\lambda(-u)\bigr)\nonumber\\
  &\times
  \mu(u)\,
  \exp{\left[i 2\pi tf (\lambda(u) - \lambda(-u))\right]}\,
  \mathrm{d}u,
\end{align}
where $p = 2r - q + 2$, with the free real parameters $r$ and $q$.
The superscript asterisk $^*$ denotes the operation of complex conjuration.
Time $t$ and frequency $f$ are real and positive real numbers correspondingly ($t\in\mathbb{R}$, $f\in\mathbb{R}^+$).
The real function $\lambda(u)$ is defined as
\begin{equation}
  \lambda (u) =
  \left(
  k \frac{\mathrm{e}^{-u} - 1}{\mathrm{e}^{-ku} - 1}
  \right)^{1/(k-1)}, \hspace{5mm}k\ne 0, 1,
\end{equation}
where $k$ is a real constant.
The function $\mu(u)$ is an arbitrary real function.

A property that makes this class interesting is its ability to provide the ideal time-frequency localization for power-law chirp signals, i.e., signals featuring a power-law functional dependence of the group delay on frequency.
Such localization is achievable for $k\le0$~\cite{bertrand1992} and with the following functional form of the group delay~\cite{bertrand1992, flandrin-geometry-1996}
\begin{equation}
    t_X(f) = t_0 + a f^{k-1},
\end{equation}
where $t_0$ and $a$ are constants.

A particular choice of $k$ selects a particular time-frequency distribution belonging to Bertrand's class.
If we would choose $k=-1$, this would correspond to the Unterberger distribution~\cite{bertrand1992} that has the function $\lambda(u)$ of the form
\begin{equation}
    \lambda(u) = \mathrm{e}^{u/2},
\end{equation}
and is able to localize the power-low chirps with the group delay~\cite{bertrand1992, flandrin-geometry-1996}
\begin{equation}
    t_X(f) = t_0 + \frac{a}{f^2}.
\end{equation}
This choice would correspond to the typical model of pulsar signals in radio astronomy~\cite{lorimer-kramer-handbook, condon-ransom-book}.

Different useful properties of Bertrand's class distributions can be achieved by choosing a specific form of $\mu(u)$.
The form required to achieve the localization property is
\begin{equation}
  \mu^{\mathrm{L}}(u) = \left| \frac{\mathrm{d}}{\mathrm{d}u}\left[ \lambda(u) - \lambda(-u) \right] \right|
  \left[\lambda(u)\lambda(-u)\right]^{r+1}.
\end{equation}
Here, the superscript $^{\mathrm{L}}$ stands for ``localized''.

Another useful property is  unitarity which should also be satisfied by the distributions.
Bertrand's class distributions feature the following general form of the unitary condition
\begin{multline}
  \int_{0}^{\infty}
  \int_{-\infty}^{\infty}
  B^{\mathrm{U} }(t,f)\,
  B^{\mathrm{U}*}(t,f)\,
  f^{2q}\,
  \mathrm{d}t\,\mathrm{d}f\\
  =
  \left|
  \int_{0}^{\infty}
  X(f)\, X^*(f)\,
  f^{2r+1}\,
  \mathrm{d}f
  \right|^2.
\end{multline}
The superscript $^{\mathrm{U}}$ stands for ``unitary''.
Unitarity can be achieved by selecting the function $\mu(u)$ in the form
\begin{equation}
    \mu^{\mathrm{U}}(u) = \sqrt{\left| \frac{\mathrm{d}}{\mathrm{d}u}\left[ \lambda(u) - \lambda(-u) \right] \right|}
  \left[\lambda(u)\lambda(-u)\right]^{r+1}.
\end{equation}

In general, Bertrand's class is not able to provide the ideal localization and unitarity simultaneously except for in the case when $k=0$.
However, if we combine the ``localized'' version of Bertrand's class with the one associated with the following ``auxiliary'' form of the $\mu(u)$ function
\begin{equation}
  \mu^{\mathrm{A}}(u) =\left[\lambda(u)\lambda(-u)\right]^{r+1},
\end{equation}
we can also fulfill the unitarity property
\begin{multline}
\label{eq:bert-prod-simple}
  \int_{0}^{\infty}
  \int_{-\infty}^{\infty}
  B^{\mathrm{L} }(t,f)\,
  B^{\mathrm{A}*}(t,f)\,
  f^{2q}\,
  \mathrm{d}t\,\mathrm{d}f\\
  =
  \left|
  \int_{0}^{\infty}
  X(f)\, X^*(f)\,
  f^{2r+1}\,
  \mathrm{d}f
  \right|^2.
\end{multline}
In this way, we relax the condition of perfect simultaneous fulfillment of perfect localization and unitarity, and instead retain the localization property for one of the distributions.
This form of the unitarity condition is central to the formulation of Bertrand's representation of the optimal detector that we will see below.

To see more details about the properties of Bertrand's class of time-frequency distributions the reader may consult references~\cite{bertrand1992} and~\cite{boashash2016}.

\section{Formulation of the detection problem}
\label{sec:formulation}
By definition~(\ref{eq:bertrand}), Bertrand's class of time-frequency distributions are computed from analytic signals, which are a type of complex signals in which the imaginary part is the Hilbert transformed real part.
For an arbitrary, discrete-time real signal
$x^{(\mathrm{r})}_{\alpha}$, the corresponding analytic signal can be formally written as~\cite{king2010, oppenheim-schafer2010}
\begin{equation}
   x_{\alpha} = x^{(\mathrm{r})}_{\alpha} + i \mathbb{H}[x^{(\mathrm{r})}_{\alpha}] = x^{(\mathrm{r})}_{\alpha} + i x^{(\mathrm{i})}_{\alpha}.
\end{equation}
Throughout this work the Greek lower index means discrete-time variable.
The upper index $^{(\mathrm{r})}$ or $^{(\mathrm{i})}$ denote real or imaginary part correspondingly.
Similar upper indices in round brackets will be used below to avoid confusion with powers.

Since our signals must be analytic we cannot simply reuse the readily available solutions for optimal detector.
The available solutions use either real or uncorrelated complex signals~\cite{flandrin1988, vantrees2001-3}.
In contrast, analytic signals are inevitably correlated due to the connection between the real and imaginary parts of signals.
Thus, first, we will develop an optimal detector for analytic signals and then we will use it to formulate Bertrand's representation.

The detection problem we want to address is the binary detection problem with two
hypotheses
\begin{equation}
\begin{aligned}
\label{eq:detection-problem}
    &H_1: r_{\alpha} = s_{\alpha} + n_{\alpha},\\
    &H_0: r_{\alpha} = n_{\alpha}.
\end{aligned}
\end{equation}
The hypothesis $H_0$ is that the observed data $r_{\alpha}$ contain only noise $n_{\alpha}$.
The opposite hypothesis $H_1$ is that the observations contain a signal of interest $s_{\alpha}$ and the additive noise.
The signals $r_{\alpha}$, $s_{\alpha}$, and $n_{\alpha}$ are the analytic signals obtained from the corresponding real signals $r^{(\mathrm{r})}_{\alpha}$, $s^{(\mathrm{r})}_{\alpha}$, and $n^{(\mathrm{r})}_{\alpha}$.

The real part of the noise $n^{(\mathrm{r})}_{\alpha}$ is assumed to be zero-mean, uncorrelated stationary Gaussian noise with the corresponding covariance matrix
\begin{equation}
\label{eq:noise-covariance}
    \mathbb{E}[n^{(\mathrm{r})}_{\alpha} n^{(\mathrm{r})}_{\beta}] = N_0 \delta_{\alpha\beta}.
\end{equation}
Here $\delta_{\alpha\beta}$ is the Kronecker delta, and $N_0$ is a constant proportional to the spectral density of the noise.

The signal of interest is assumed to be such that its real part, $s^{(\mathrm{r})}_{\alpha}$, belongs to the class of Gaussian signals with the given expected value
\begin{equation}
\label{eq:initialmean}
    \mathbb{E}[s^{(\mathrm{r})}_{\alpha}] = m^{(\mathrm{r})}_{\alpha},
\end{equation}
and covariance matrix
\begin{equation}
\label{eq:initialcov}
    \mathbb{E}[(s^{(\mathrm{r})}_{\alpha} - m^{(\mathrm{r})}_{\alpha})(s^{(\mathrm{r})}_{\beta} - m^{(\mathrm{r})}_{\beta})] = K_{\alpha\beta}.
\end{equation}

The remaining correlation properties of all other components of the analytic signals emerge from the ones available for the real parts of the signals (see Appendix~\ref{app:covariances} for details).

\section{General case of the optimal detector of complex signals}
\label{sec:opt-det-general}
The fundamental component in building the optimal detector is the selection of the appropriate distribution of the pure noise~\cite{shirman1981}.
Since the structure of the correlation matrix of the analytic signal is apriori unknown, we use a generalization of the multivariate complex normal distribution~\cite{bos1995} to avoid possible limitations of simpler distributions.
This generalization allows the use of an arbitrary covariance matrix to incorporate the desired structure of the correlations.
This distribution will serve as the basis for the optimal detector in this work.

Let $z_1$, $z_2, \dots, z_N$ be the complex random variables associated to the samples of an analytic signal $z_{\alpha}$.
These variables and their complex conjugates are arranged in a vector as follows
\begin{equation}
\label{eq:vvec}
  \mathbf{v} = (z_1, z_1^*, \dots, z_N, z_N^*)^{\intercal}.
\end{equation}
The superscript $^{\intercal}$ denotes the operation of transposition.
The covariance matrix associated with these variables is defined as the expectation value of the vector $\mathbf{v}$ multiplied to its hermitian conjugate version $\mathbf{v}^{\dag}$
\begin{equation}
  V = \mathbb{E}[\mathbf{v}\mathbf{v}^{\dag}].
\end{equation}
The multivariate complex normal distribution that we use is defined as~\cite{bos1995}
\begin{equation}
  \mathcal{N}(\mathbf{v}) = \frac{1}{\pi^N (\mathrm{det}V)^{1/2}}
  \exp \left(-\frac{1}{2} \mathbf{v}^{\dag} V^{-1} \mathbf{v}\right),
  \label{eq:complexnormal}
\end{equation}
where $\mathrm{det}V$ is the determinant of $V$ and $V^{-1}$ is the inverse matrix to $V$.

Now, we follow the procedure in \cite{shirman1981} and build the detector based on the logarithm of the likelihood ratio, $\Lambda$, (hereafter log-likelihood ratio) of the likelihood functions corresponding to the distributions with signal and additive noise $\mathcal{L}_{H_1}$ and with the noise only $\mathcal{L}_{H_0}$.

We use (\ref{eq:complexnormal}) as the corresponding probability density function and consider all the signals $r_{\alpha}$, $s_{\alpha}$, and $n_{\alpha}$ as the corresponding vectors.
The presence of the signal causes a shift of the mean value of the distribution to the expected value of the signal, and the covariance matrix has the corresponding contribution from the presence of the signal
\begin{equation}
  \mathcal{L}_{H_1} = \xi_{H_1}
  \exp
  \left(-\frac{1}{2}
  (\mathbf{r} - \mathbf{m})^{\dag} V_{H_1}^{-1} (\mathbf{r} - \mathbf{m})
  \right).
\end{equation}
Without the signal, the likelihood function is simply
\begin{equation}
  \mathcal{L}_{H_0} = \xi_{H_0}
  \exp
  \left(-\frac{1}{2}
  \mathbf{r}^{\dag} V_{H_0}^{-1} \mathbf{r}
  \right).
\end{equation}
The letters $\xi_{H_1}$ and $\xi_{H_0}$ denote the normalization constants that are not relevant for the final form of the detector.

The log-likelihood ratio can be written straightforwardly
\begin{multline}
\label{eq:lnl-preform}
\ln \Lambda = \ln\mathcal{L}_{H_1} - \ln\mathcal{L}_{H_0} = \\
-\frac{1}{2} (\mathbf{r} - \mathbf{m})^{\dag} V_{H_1}^{-1} (\mathbf{r} - \mathbf{m})
+\frac{1}{2}
  \mathbf{r}^{\dag} V_{H_0}^{-1} \mathbf{r}
  + C.
\end{multline}
The constant $C$ contains all the components of $\ln \Lambda$ that do not depend on the observed signal.

By expanding (\ref{eq:lnl-preform}) and hiding the term $\mathbf{m}^{\dag} V_{H_1}^{-1} \mathbf{m}$ in the constant $C$, we arrive at the general form of the optimal detector for complex signals with arbitrary covariance matrix
\begin{multline}
\label{eq:lnl}
\ln\Lambda =
\frac{1}{2}\left[
- \mathbf{r}^{\dag} V_{H_1}^{-1} \mathbf{r}
+ \mathbf{r}^{\dag} V_{H_0}^{-1} \mathbf{r}
\right] \\
+ \frac{1}{2}\left[
\mathbf{m}^{\dag} V_{H_1}^{-1} \mathbf{r}
+ \mathbf{r}^{\dag} V_{H_1}^{-1} \mathbf{m}
\right] + C.
\end{multline}
The first and second brackets can be considered to represent the detector components associated with the random and deterministic aspects of the signal, respectively.

\section{Optimal detector of analytic signals}
\label{sec:opt-det-analytical}
To obtain the concrete form of the optimal detector for analytic signals, we should investigate the structure of the corresponding covariance matrices.
We devote Appendix~\ref{app:cov-matrix-structure} to this study.

The structure of the correlation matrix for the analytic signal is such that we can equivalently rewrite the detector (\ref{eq:lnl}) in a simpler form.
Instead of the vectors (\ref{eq:vvec}), we can use vectors of a simpler structure
\begin{equation}
    \mathbf{u} = (z_1 \dots z_N)^{\intercal}.
\end{equation}
and the associated covariance matrix
\begin{equation}
    U = \mathbb{E}[\mathbf{u}\mathbf{u}^{\dag}]
\end{equation}
to obtain the equivalent form of the detector (\ref{eq:lnl})
\begin{multline}
\label{eq:detector-prefinal-form}
\ln\Lambda =
\frac{1}{2}\left[
- \mathbf{r}^{\dag} U_{H_1}^{-1} \mathbf{r}
+ \mathbf{r}^{\dag} U_{H_0}^{-1} \mathbf{r}
\right] \\
+ \frac{1}{2}\left[
\mathbf{m}^{\dag} U_{H_1}^{-1} \mathbf{r}
+ \mathbf{r}^{\dag} U_{H_1}^{-1} \mathbf{m}
\right] + c.c. + C.
\end{multline}
The notation $c.c.$ stands for ``complex conjugate'' and denotes the complex conjugate of all the terms in front.

The covariance matrices in (\ref{eq:detector-prefinal-form}) can be decomposed to the Karhunen-Lo\`{e}ve series (see Appendix \ref{app:kl} for details)
\begin{align}
\label{eq:KL-decomposition}
    U_{H_1}^{-1} &= \sum_{i=1}^{N} \frac{1}{2 N_0 + \eta_i^2} \boldsymbol{\phi}_i \boldsymbol{\phi}_i^{\dag},\\
    U_{H_0}^{-1} &= \sum_{i=1}^{N} \frac{1}{2 N_0} \boldsymbol{\phi}_i \boldsymbol{\phi}_i^{\dag},
\end{align}
where $\boldsymbol{\phi}_i$ and $\eta_i^2$ are eigenvectors and eigenvalues of the decomposition.
On the one hand, this decomposition is part of the classical solution~\cite{flandrin1988, vantrees2001-3} and should be done for completeness.
On the other hand, this decomposition is essential for obtaining pairs of internal products, which are then required for representing the detector in terms of Bertrand's class.
The situation is analogous to that observed in the case of Cohen's class representation of the optimal detector~\cite{flandrin1988}.

The  Karhunen-Lo\`{e}ve decomposition of (\ref{eq:detector-prefinal-form}) yields the following result
\begin{equation}
\label{eq:lnl-prefinal}
    \ln\Lambda = \frac{1}{2} l_{\mathrm{R}} + \frac{1}{2} l_{\mathrm{D}} + c.c. + C,
\end{equation}
where
\begin{align}
\label{eq:lr-initial}
    l_{\mathrm{R}} &=
    - \sum_{i=1}^{N} \frac{1}{2 N_0 + \eta_i^2}
    \mathbf{r}^{\dag} \boldsymbol{\phi}_i \boldsymbol{\phi}_i^{\dag} \mathbf{r}
    + \sum_{i=1}^{N} \frac{1}{2 N_0}
    \mathbf{r}^{\dag} \boldsymbol{\phi}_i \boldsymbol{\phi}_i^{\dag} \mathbf{r},\\
\label{eq:ld-initial}
    l_{\mathrm{D}} &=
    \sum_{i=1}^{N} \frac{1}{2 N_0 + \eta_i^2}
    \mathbf{m}^{\dag} \boldsymbol{\phi}_i \boldsymbol{\phi}_i^{\dag} \mathbf{r}
    + \sum_{i=1}^{N} \frac{1}{2 N_0 + \eta_i^2}
    \mathbf{r}^{\dag} \boldsymbol{\phi}_i \boldsymbol{\phi}_i^{\dag} \mathbf{m}.
\end{align}
The subscripts $_{\mathrm{R}}$ and $_{\mathrm{D}}$ stand for ``random'' and ``deterministic'' following the nature of the terms.

By considering the pairs of internal products, the terms of the detector can be further simplified by using the fact that the internal product values remain constant after a transpose operation, as they are scalars (see Appendix \ref{app:simplification} for details)
\begin{align}
\label{eq:lr-after-simplification}
    l_{\mathrm{R}} &= \frac{1}{2 N_0} \sum_{i=1}^{N} \frac{\eta_i^2}{2 N_0 + \eta_i^2}
    |\boldsymbol{\phi}_i^{\dag} \mathbf{r} |^2,\\
\label{eq:ld-after-simplification}
    l_{\mathrm{D}} &=
    2 \sum_{i=1}^{N} \frac{1}{2 N_0 + \eta_i^2}
    (\mathbf{m}^{\dag} \boldsymbol{\phi}_i)(\boldsymbol{\phi}_i^{\dag} \mathbf{r}).
\end{align}

Finally, we can consider the complex conjugate part of (\ref{eq:lnl-prefinal}) together with the main part.
If we do so and use the basic property $z + z^* = 2\mathrm{Re}\{z\}$, we arrive at the following form of the detector
\begin{multline}
\label{eq:lnl-prefinal2}
    \ln\Lambda =
    \frac{1}{2 N_0} \sum_{i=1}^{N} \frac{\eta_i^2}{2 N_0 + \eta_i^2}
    |\boldsymbol{\phi}_i^{\dag} \mathbf{r} |^2\\
    + 2 \mathrm{Re}\left\{ \sum_{i=1}^{N} \frac{1}{2 N_0 + \eta_i^2}
    (\mathbf{m}^{\dag} \boldsymbol{\phi}_i)(\boldsymbol{\phi}_i^{\dag} \mathbf{r})\right\} + C.
\end{multline}
The internal products that appeared here can be seen as the corresponding integrals
\begin{equation}
    |\boldsymbol{\phi}_i^{\dag} \mathbf{r}|^2 =
    \left|\int_{-T}^{T} r(t) \phi_i^*(t) \, \mathrm{d}t\right|^2.
\end{equation}
\begin{multline}
    (\mathbf{m}^{\dag} \boldsymbol{\phi}_i)(\boldsymbol{\phi}_i^{\dag} \mathbf{r})
    = (\mathbf{m}^{\intercal} \boldsymbol{\phi}_i^*)^*(\boldsymbol{\phi}_i^{\dag} \mathbf{r})\\
    = \left[ \int_{-T}^{T} m(t) \phi_i^*(t)\, \mathrm{d}t \right]^*
    \left[ \int_{-T}^{T} r(t) \phi_i^*(t)\, \mathrm{d}t \right]
\end{multline}
Even though this notation is convenient for further use, it is rather
formal.

By inserting this integral form of the internal products to (\ref{eq:lnl-prefinal2}) we arrive at the final form of the optimal detector of analytic signals in analytic noise
\begin{equation}
\label{eq:lnl-final}
    \ln\Lambda = l_{\mathrm{R}} + l_{\mathrm{D}},
\end{equation}
where
\begin{equation}
\label{eq:lr-final}
    l_{\mathrm{R}} = \frac{1}{2 N_0} \sum_{i=1}^{N} \frac{\eta_i^2}{2 N_0 + \eta_i^2}
    \left|\int_{-T}^{T} r(t) \phi_i^*(t) \, \mathrm{d}t\right|^2,
\end{equation}
\begin{multline}
\label{eq:ld-final}
    l_{\mathrm{D}} =2 \mathrm{Re}\left\{ \sum_{i=1}^{N} \frac{1}{2 N_0 + \eta_i^2}\right.
    \left[ \int_{-T}^{T} r(t) \phi_i^*(t)\, \mathrm{d}t \right] \\
    \times \left. \left[ \int_{-T}^{T} m(t) \phi_i^*(t)\, \mathrm{d}t \right]^*\right\},
\end{multline}
and the constant $C$ from (\ref{eq:lnl-prefinal2}) can be incorporated to the detection threshold since it does not change the functional part of the detector.

The detection procedure is
\begin{equation}
\label{eq:detection-procedure}
  \ln\Lambda \overset{H_1}{\underset{H_0}{\gtrless}} \gamma.
\end{equation}
The letter $\gamma$ denotes the desired detection threshold that can be set by using the Neyman-Pearson procedure to stabilize the rate of the false detections~\cite{vantrees2001-3}.

The detector (\ref{eq:lnl-final}), (\ref{eq:lr-final}), (\ref{eq:ld-final}) and the detection procedure (\ref{eq:detection-procedure}) is the solution to the problem (\ref{eq:detection-problem}).

\section{Bertrand's representation of the optimal detector}
\label{sec:opt-det-bert}
The main structural element of the obtained detector components (\ref{eq:lr-final}) and (\ref{eq:ld-final}) is the combination of the internal products.
This can be written in the general way
\begin{equation}
  \left[
  \int_{(T)}
  x_1(t)\, x^*_2(t)
  \mathrm{d}t
  \right]
  \left[
  \int_{(T)}
  x_3(t)\, x^*_4(t)
  \mathrm{d}t
  \right]^*  
\end{equation}
where functions $x_1(t)$, $x_2(t)$, $x_3(t)$, and $x_4(t)$ correspond the certain functions in the particular cases.

To find Bertrand's class representation of the optimal detector, we use the same idea as in~\cite{flandrin1988}.
Namely, we rewrite the pairs of internal products in terms of Bertrand's class time-frequency distributions.

Firstly, we expand the conventional definition of Bertrand's class (\ref{eq:bertrand}) to the cross-Bertrand's class, which we define as follows
\begin{align}
  B_{XY}(t, f) =
  f^p
  &\int_{-\infty}^{\infty}
  X\bigl(f\lambda(u)\bigr)\,
  Y^*\bigl(f\lambda(-u)\bigr)\nonumber\\
  &\times
  \mu(u)\,
  \exp{\left[i 2\pi tf (\lambda(u) - \lambda(-u))\right]}\,
  \mathrm{d}u.
\end{align}
The only difference from the original definition (\ref{eq:bertrand}) is consideration of the two different signals $X(f)$ and $Y(f)$ instead of one.
The functions $\lambda(u)$ and $\mu(u)$, and the constant $p$ are the same.

If we use the unitary property for the original Bertrand's class (\ref{eq:bert-prod-simple}) as a guide, a straightforward computation leads to the following relation for the cross-Bertrand's classes for the four signals $X_1(f)$, $X_2(f)$, $X_3(f)$, and $X_4(f)$
\begin{multline}
\label{eq:bert-prod-general}
  \int_{0}^{\infty}
  \int_{-\infty}^{\infty}
  B_{X_1 X_2}^{\mathrm{L} }(t,f)\,
  B_{X_3 X_4}^{\mathrm{A}*}(t,f)\,
  f^{2q}\,
  \mathrm{d}t\,\mathrm{d}f\\
  =
  \left[
  \int_{0}^{\infty}
  X_1(f)\, X^*_3(f)\,
  f^{2r+1}\,
  \mathrm{d}f
  \right] \\
  \times
  \left[
  \int_{0}^{\infty}
  X_2(f)\, X^*_4(f) \,
  f^{2r+1}\,
  \mathrm{d}f
  \right]^*.
\end{multline}
We can observe the change of the original sequence of signals from $\{x_1, x_2, x_3, x_4\}$ to $\{x_1, x_3, x_2, x_4\}$ where $x_2$ and $x_3$ are permuted.
The same change happens for a similar relationship relevant for Cohen's class~\cite{flandrin1988}.

By selecting the free parameters as $r=1/2$ and $q=0$, we obtain the desired connection between the pair of internal products and Bertrand's class
\begin{multline}
\label{eq:bert-prod}
  \int_{0}^{\infty}
  \int_{-\infty}^{\infty}
  B_{X_1 X_2}^{\mathrm{L} }(t,f)\,
  B_{X_3 X_4}^{\mathrm{A}*}(t,f)\,
  \mathrm{d}t\,\mathrm{d}f\\
  =
  \left[
  \int_{0}^{\infty}
  X_1(f)\, X^*_3(f)
  \mathrm{d}f
  \right]
  \left[
  \int_{0}^{\infty}
  X_2(f)\, X^*_4(f)
  \mathrm{d}f
  \right]^*.
\end{multline}
The signals can be equivalently represented in the time domain by applying Parseval's theorem
\begin{multline}
\label{eq:bert-prod2}
  \int_{0}^{\infty}
  \int_{-\infty}^{\infty}
  B_{X_1 X_2}^{\mathrm{L} }(t,f)\,
  B_{X_3 X_4}^{\mathrm{A}*}(t,f)\,
  \mathrm{d}t\,\mathrm{d}f\\
  =
  \left[
  \int_{(T)}
  x_1(t)\, x^*_2(t)
  \mathrm{d}t
  \right]
  \left[
  \int_{(T)}
  x_3(t)\, x^*_4(t)
  \mathrm{d}t
  \right]^*.
\end{multline}
Here, $x_i(t)$ are corresponding time-domain representations of the analytic signals $X_i(f)$.

The last equation applied to the equations (\ref{eq:lr-final}) and (\ref{eq:ld-final}) yields Bertrand's representation of the optimal detector
\begin{multline}
\label{eq:lr-bert}
    l_{\mathrm{R}} = \frac{1}{2 N_0} \sum_{i=1}^{N} \frac{\eta_i^2}{2 N_0 + \eta_i^2}\\
    \times
    \int_{0}^{\infty}
    \int_{-\infty}^{\infty}
    B_{rr}^{\mathrm{L} }(t,f)\,
    B_{\phi_i \phi_i}^{\mathrm{A}*}(t,f)\,
    \mathrm{d}t\,\mathrm{d}f,
\end{multline}
\begin{multline}
\label{eq:ld-bert}
    l_{\mathrm{D}} =2 \mathrm{Re}\Biggl\{ \sum_{i=1}^{N} \frac{1}{2 N_0 + \eta_i^2}\\
    \times
    \int_{0}^{\infty}
    \int_{-\infty}^{\infty}
    B_{rm}^{\mathrm{L} }(t,f)\,
    B_{\phi_i \phi_i}^{\mathrm{A}*}(t,f)\,
    \mathrm{d}t\,\mathrm{d}f
    \Biggr\}.
\end{multline}

The obtained form is not unique.
The functions $\mu^{\mathrm{L}}(u)$ and $\mu^{\mathrm{A}}(u)$ in Bertrand's classes in (\ref{eq:bert-prod-general}) can be swapped and the combination
$B_{X_1 X_2}^{\mathrm{A} }(t,f)\, B_{X_3 X_4}^{\mathrm{L}*}(t,f)$
can be used instead of
$B_{X_1 X_2}^{\mathrm{L} }(t,f)\, B_{X_3 X_4}^{\mathrm{A}*}(t,f)$.
The resulting equation is going to be identical
\begin{multline}
  \int_{0}^{\infty}
  \int_{-\infty}^{\infty}
  B_{X_1 X_2}^{\mathrm{A} }(t,f)\,
  B_{X_3 X_4}^{\mathrm{L}*}(t,f)\,
  \mathrm{d}t\,\mathrm{d}f\\
  =
  \left[
  \int_{(T)}
  x_1(t)\, x^*_3(t)
  \mathrm{d}t
  \right]
  \left[
  \int_{(T)}
  x_2(t)\, x^*_4(t)
  \mathrm{d}t
  \right]^*.
\end{multline}
Thus, the classes $B_{x y}^{\mathrm{L} }$ and $B_{x y}^{\mathrm{A} }$ in (\ref{eq:lr-bert}) and in (\ref{eq:ld-bert}) can be swapped leading to an equivalent formulation of the optimal detector.

We can yet obtain another, symmetric form of the optimal detector by selecting the function $\mu(u)$ in the form
\begin{equation}
    \mu^{\mathrm{U}}(u) = \sqrt{\left| \frac{\mathrm{d}}{\mathrm{d}u}\left[ \lambda(u) - \lambda(-u) \right] \right|}
  \left[\lambda(u)\lambda(-u)\right]^{r+1}.
\end{equation}
The corresponding Bertrand's classes are unitary with themselves
\begin{multline}
  \int_{0}^{\infty}
  \int_{-\infty}^{\infty}
  B_{X_1 X_2}^{\mathrm{U} }(t,f)\,
  B_{X_3 X_4}^{\mathrm{U}*}(t,f)\,
  \mathrm{d}t\,\mathrm{d}f\\
  =
  \left[
  \int_{(T)}
  x_1(t)\, x^*_3(t)
  \mathrm{d}t
  \right]
  \left[
  \int_{(T)}
  x_2(t)\, x^*_4(t)
  \mathrm{d}t
  \right]^*,
\end{multline}
which means that one more form of (\ref{eq:lr-bert}) and (\ref{eq:ld-bert}) can be readily obtained by using $B_{x y}^{\mathrm{U} }$ on the places of $B_{x y}^{\mathrm{L} }$ and $B_{x y}^{\mathrm{A} }$.

\section{Examples}
\label{sec:examples}
This section presents two examples to illustrate the equations (\ref{eq:lr-bert}) and (\ref{eq:ld-bert}) for particular cases.
The examples are the same as in~\cite{flandrin1988}.

\subsection{Fully known signal}
Let us assume that the analytic signal of interest, is completely described by the function $g_{\alpha}$
\begin{equation}
    s_{\alpha}= g_{\alpha}.
\end{equation}
This means that its expected value (\ref{eq:initialmean}) is the known function
\begin{equation}
    m_{\alpha} = g_{\alpha},
\end{equation}
and the covariance matrix (\ref{eq:initialcov}) is reduced to zero
\begin{equation}
    K_{\alpha\beta} = 0.
\end{equation}

In this case the term (\ref{eq:lr-bert}) associated to the random part of the signal of interest is zero
\begin{equation}
    l_{\mathrm{R}} = 0,
\end{equation}
but the term (\ref{eq:ld-bert}) associated to the deterministic part is non-zero.
It has the form of the time-frequency correlation of the cross-Bernard's class distribution between the observed and the expected signal and Bertrand's class distribution of the expected signal.
Namely
\begin{multline}
    l_{\mathrm{D}} = 2 \mathrm{Re}\Biggl\{ \frac{1}{2 N_0 + \eta_g^2} \\
    \times\int_{0}^{\infty}
    \int_{-\infty}^{\infty}
    B_{rg}^{\mathrm{L} }(t,f)\,
    B_{gg}^{\mathrm{A}*}(t,f)\,
    \mathrm{d}t\,\mathrm{d}f
    \Biggr\},
\end{multline}
where the constant $\eta^2_g$ is equal to the energy $E_g$ of the expected signal.
This form of the detector generalizes the case in~\cite{chassande-mottin1999}, where the correlation is performed with the ideal power-law chirp signal to eventually simplify the detection to the path integration on the time-frequency plane.
The present version of the detector extends this idea to any power-law chirp signal and enables its optimal detection.

\subsection{Rayleigh fading signal}
In this example the signal of interest is the known analytic signal with unknown, random amplitude $b$
\begin{equation}
    s_{\alpha} = b g_{\alpha}.
\end{equation}
The amplitude $b$ is assumed to be a real Gaussian random variable, such that
\begin{align}
    \mathbb{E}[b] &=0,\\
    \mathbb{E}[b^2] &= \sigma_b ^2.
\end{align}
In this case, the mean expected signal (\ref{eq:initialmean}) is zero
\begin{equation}
    m_{\alpha} = 0,
\end{equation}
and the covariance matrix (\ref{eq:initialcov}) can be expressed as
\begin{equation}
    K_{\alpha\beta} = \sigma_b^2\, g_{\alpha} g^*_{\beta}.
\end{equation}
The corresponding eigenvalue is
\begin{equation}
    \eta_g^2 = \sigma_b^2 E_g,
\end{equation}
where $E_g$ is the energy of the signal $g_{\alpha}$.

In this case, opposite to the previous example, the part associated to the deterministic part of the detector is zero
\begin{equation}
    l_{\mathrm{D}} =0,
\end{equation}
and the random part is non-trivial
\begin{equation}
    l_{\mathrm{R}} = \frac{1}{2 N_0} \frac{\eta_g^2}{2 N_0 + \eta_g^2}
    \int_{0}^{\infty}
    \int_{-\infty}^{\infty}
    B_{rr}^{\mathrm{L} }(t,f)\,
    B_{gg}^{\mathrm{A}*}(t,f)\,
    \mathrm{d}t\,\mathrm{d}f.
\end{equation}
As in the previous example, the detector has the form of the time-frequency correlation, but now of Bertrand's class distributions of the observed and expected signals.

\section{Example numerical simulation}
The present paper summarizes the theoretical development of the new time-frequency representation of the optimal detector for analytic signals.
One of the naturally arising questions is whether such a representation is useful for some situations.
To address this question and also to fully assess the performance of the representation, one needs to perform a comprehensive simulation study.
Such a study is out of the scope of this paper.
However, a particular case where this representation seems useful can be presented already here already.

Since the prime interest to Bertrand's representation arose in the context of radio astronomy, we consider here a signal with the group delay of the form~(\ref{eq:group-delay}) which is relevant for detection of the pulsar and fast radio burst signals in radio astronomy.
Namely, we consider the following model of a deterministic discrete analytic signal
\begin{equation}
    g_{\alpha} = A_g \left( \cos \sqrt{b \alpha} + i\mathbb{H}[\cos \sqrt{b \alpha}] \right) w_{\alpha},
    \label{eq:test-chirp}
\end{equation}
where $b$ is a real parameter (here $b=4\times10^5\times(2\pi)^2$), $\alpha$ is the discrete time parameter following the notation we introduced above ($\alpha \in \mathbb{Z}$), and $A_g$ is the normalization factor which we choose to normalize the signal energy to unity.
The signal consists 1024 samples ($\alpha=0, 1,\dots,1023$) and windowed with a rectangular window $w_{\alpha}$ such that the first and the last 250 samples are set to zero.
We consider an analytic signal here since the Bertrand's class distribution by construction operate only with this type of signals.

In this simulations, we use the additive noise, $n_{\alpha}$, which is discrete and analytic.
Its real part is the uncorrelated white noise such that $N_0=1$ (see (\ref{eq:noise-covariance})).

The resulting signal analyzed by the detector is
\begin{equation}
    r_{\alpha} = \sqrt{E}\,g_{\alpha} + n_{\alpha},
\end{equation}
where $E$ denotes the energy of the observed signal $g_{\alpha}$.

Firstly, we simulate three representation of the optimal detector for a fully known signal: time-domain detector (matched filter), its Wigner-Ville's representation, and its Bertrand's representation.
We use the discrete versions of all distributions (denoted here with tilde, $\tilde{\cdot}$, for clarity) and the corresponding discrete formulas for the detectors.
We ignore the normalization of the detectors since it only scales the detection threshold which is determined directly in the simulation for each of the detectors.
Thus, the detectors are reduced to their functional form only.
The time-domain optimal detector~(\ref{eq:lnl-final}) becomes
\begin{equation}
    l =
    \mathrm{Re}\left\{
    \sum_{n=0}^{N} r(n) g^*(n)
    \right\}.
    \label{eq:test-det-mf}
\end{equation}

For the Wigner-Ville's representation of the optimal detector we use
\begin{equation}
    l =
    \mathrm{Re}\left\{
    \sum_{n=0}^{N}
    \sum_{m=0}^{M}
    \tilde{W}_{rg}(n,m) \tilde{W}_{gg}^*(n,m)
    \right\}
    \label{eq:test-det-wv}
\end{equation}
with the discrete implementation from~\cite{chassande-mottinDiscreteTimeFrequency2005} to ensure conservation of the Moyal relation.

As an example of the Bertrand's class detector, we use the detector based on the Unterberger distribution (corresponds to $k=-1$ of the class and enables the perfect localization of the considered signal)
\begin{equation}
    l =
    \mathrm{Re} \left\{
    \sum_{n=0}^{N}
    \sum_{m=0}^{M}
    \tilde{B}^{\mathrm{L}}_{rg}(n,m) \tilde{B}^{\mathrm{A}*}_{gg}(n,m)
    \right\}.
    \label{eq:test-det-ub}
\end{equation}

For all these cases, the random part of the detector is always zero ($l_R=0$).

\begin{figure}[!t]
\centering
\includegraphics[width=1.\linewidth]{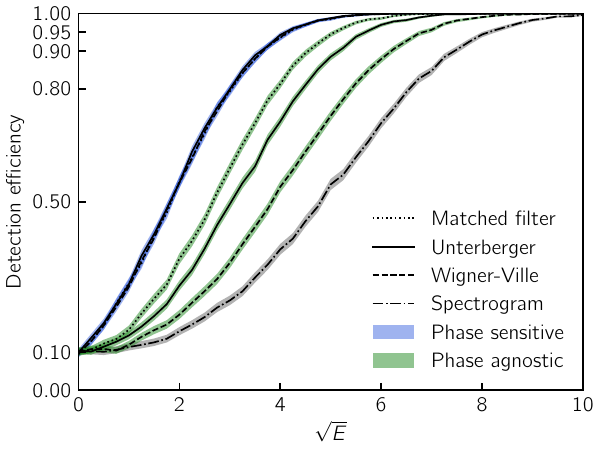}
\caption{
Simulated detection efficiencies are shown for three different optimal detectors: time-domain detector (matched filter), its Wigner-Ville's representation, and its Bertrand's representation.
All the detectors were considered in the phase-sensitive and phase-agnostic forms.
For comparison purposes, we also present the detection efficiency of the spectrogram correlation-based detector.
The lines connect the points at which the computations were performed. The bands indicate the 68\% confidence interval.
The zigzag shape of the curves reflects the probabilistic nature of the underlying computations.
The computations were performed with a step size of 0.25 on the horizontal axis.}
\label{fig-sens}
\end{figure}

We assess the performance of the detectors by evaluating their efficiency to detect the signal~(\ref{eq:test-chirp}) of a given energy.
The value of the detection efficiency at a given energy we estimate as the fraction of times a given value $l$ exceeds the predefined threshold in 10$^4$ trials.
The threshold in this procedure is defined with the Neyman-Pearson criterion with 10\% of the false positive rate for presentation purposes.
The uncertainties associated with the estimated efficiency values were computed as the Wilson intervals~\cite{brown_interval_2001}.
Fig.~\ref{fig-sens} presents the resulting detection efficiencies.

Secondly, let us consider a more realistic type of detectors.
Those that are insensitive to the constant phase offset of the signal.
To obtain such detectors, we substitute the real part operator with the absolute value operator ($\mathrm{Re}\{~\cdot~\}$ with $|~\cdot~|$) in the detectors for fully known signals~(\ref{eq:test-det-mf}), (\ref{eq:test-det-mf}), and (\ref{eq:test-det-ub})~\cite{helstromStatisticalTheorySignal1968}.
For these phase-insensitive detectors we follow the same procedure as before to estimate their detection efficiencies and depict them on the same plot in Fig.~\ref{fig-sens}.

For comparison, we also compute the detection efficiency of the detector based on the spectrogram correlation.
This method is similar to the one used in modern radio astronomy for the detection of pulsars.
Namely, if $\tilde{S}_x$ is the discrete spectrogram for a given discrete signal $x_{\alpha}$, then we can consider the following detector
\begin{equation}
    l = \sum_{n=0}^{N}
    \sum_{m=0}^{M}
    \tilde{S}_r(n,m) \tilde{S}_g(n,m).
\end{equation}
Following the same procedure as for all other detectors, we compute the detection efficiency (see Fig.~\ref{fig-sens}).
As expected from the theory~\cite{flandrin1988}, the spectrogram correlation has suppressed efficiency in comparison to the time-domain optimal detector.

The obtained detection efficiencies reveal two effects:
\begin{enumerate}
    \item all representations of the optimal detector for the fully known signals show the identical performance,
    \item Bertrand's representation of the optimal detector performs is superior among the time-frequency representations for the case of the phase agnostic detection of power-law chirps.
\end{enumerate}
The first effect is expected from the theory and indicates that all the representations are obtained correctly.
The later effect highlights one of the situations where the new type of the time-frequency representation of the optimal detectors can be useful, namely, detection of the power-law chirp signals with unknown phase.
The nature of the latter effect requires additional investigation. 
However, in this work it appears to be associated with the fact that the distribution used to form the detector is able to localize the signal in an ideal manner.

\section{Conclusion}
The paper presents the representation of the optimal detector of analytic signals in terms of Bertrand's class of time-frequency distributions.
The main results are the equations (\ref{eq:lnl-final}), (\ref{eq:detection-procedure}), (\ref{eq:lr-bert}), and (\ref{eq:ld-bert}).

This representation extends the ideas expressed in~\cite{flandrin1988, chassande-mottin1999} of the time-frequency formulation of the optimal detection to the realm of Bertrand's class which is more relevant to the power-law chirp signals due to the construction of the class.

To illustrate the performance of the new framework, we compared its detection efficiency against known methods (the matched filter, Wigner-Ville detector, and spectrogram correlation detector) for the cases of a fully known signal and a signal with the unknown phase.
The results clearly demonstrate the superiority of the new method over the other time-frequency representations.

The effects observed in the simulations and the properties of the new representation are the subject of a separate detailed investigation.

\section*{Acknowledgment}
The author would like to thank Dominik J.~Schwarz for discussing the matters described in this paper and for helping with its preparation,
Andrei Kazantsev for detailed discussions of methods used in the analysis of pulsar signals,
Dmitriy Kostunin for valuable suggestions on the improvement of the organization of the paper, and Travis Dore for proofreading the final manuscript.

Funded by the Deutsche Forschungsgemeinschaft (DFG, German Research Foundation) – project number 460248186 (PUNCH4NFDI).

{\appendices
\section{Correlation properties of analytic signal}
\label{app:covariances}
In the formulation of detection problem in Section~\ref{sec:formulation}, we made assumptions regarding the type of the real signals for which the analytic signal should be constructed and, in turn, detected.
The covariance properties of the real signals are assumed to be known.
In this appendix, we will derive all other covariance properties and see that they can be expressed from the known covariance of the real parts of the analytic signal.
We use here the usual notation of a complex number $z = x + i y$ to simplify the notation.
For completeness, we perform all computations in detail.

The logic of the derivation follows the one used in~\cite{vainshtein1983} but here the derivation is done for the case of discrete-time signals.

The basic equation that we use is the covariance function of the real part of the analytic signal
\begin{equation}
\label{eq:Rxx}
    C_{\beta}^{(xx)} = \mathbb{E}\left[x_{\alpha} x_{\alpha-\beta}\right].
\end{equation}
The symmetry property ($C^{(xx)}_{\beta} = C^{(xx)}_{-\beta}$) holds in this case since we assume that the process is stationary in at least the wide sense~\cite{davenport-root, therrien}.
In this notation, the correlation of the imaginary and real parts of the signal is
\begin{multline}
\label{app-eq:Cyx-init}
    C^{(yx)}_{\beta}
    = \mathbb{E} \left[x_{\alpha-\beta} \sum_{\gamma = -\infty}^{\infty} h_{\gamma} x_{\alpha-\gamma}\right] \\
    = \sum_{\gamma = -\infty}^{\infty} h_{\gamma} \mathbb{E}\left[x_{\alpha-\beta} x_{\alpha-\gamma}\right].
\end{multline}
Here, we expressed the imaginary part of the analytic signal by its definition via the real part of the signal
\begin{equation}
    y_{\alpha} = \sum_{\gamma = -\infty}^{\infty} h_{\gamma} x_{\alpha-\gamma},
\end{equation}
where $h_{\gamma}$ are the corresponding coefficients of the discrete Hilbert transform~\cite{king2010}
\begin{equation}
    h_{\gamma} =
    \begin{dcases}
        0, & \gamma=0, \\
        \frac{2 \sin^2 (\pi \gamma / 2)}{\pi \gamma}, & \gamma\ne0.
    \end{dcases}
\end{equation}

We can change the variables $\alpha' = \alpha-\beta$ in (\ref{app-eq:Cyx-init}) for simplification
\begin{equation}
    \sum_{\gamma = -\infty}^{\infty} h_{\gamma} \mathbb{E}\left[x_{\alpha'} x_{\alpha'+\beta-\gamma}\right]
    = \sum_{\gamma = -\infty}^{\infty} h_{\gamma} C^{(xx)}_{\beta-\gamma}.
\end{equation}
The resulting equation is the explicit form of the discrete Hilbert transform, which brings us to the first result
\begin{equation}
\label{eq:c_yx}
     C^{(yx)}_{\beta} = \mathbb{H}[C^{(xx)}_{\beta}].
\end{equation}

To find the correlation of the real to imaginary parts, we use the same logic and in addition the symmetry property $C^{(xx)}_{\beta} = C^{(xx)}_{-\beta}$.
Namely
\begin{multline}
    C^{(xy)}_{\beta}
    = \mathbb{E} \left[x_{\alpha} \sum_{\gamma = -\infty}^{\infty} h_{\gamma} x_{\alpha-\beta-\gamma}\right] = \\
    = \sum_{\gamma = -\infty}^{\infty} h_{\gamma} C^{(xx)}_{-\beta-\gamma}
    = \sum_{\gamma = -\infty}^{\infty} h_{\gamma} C^{(xx)}_{\beta+\gamma}.
\end{multline}
Now, we can change the variable $\gamma$ to $-\gamma'$ and use the property $h_{-\gamma} = -h_{\gamma}$ to find the form resembling correlation
\begin{equation}
    \sum_{\gamma' = -\infty}^{\infty} h_{-\gamma'} C^{(xx)}_{\beta-\gamma'}
    = - \sum_{\gamma' = -\infty}^{\infty} h_{\gamma'} C^{(xx)}_{\beta-\gamma'}.
\end{equation}
The resulting equation gives the desired result
\begin{equation}
\label{eq:c_xy}
    C^{(xy)}_{\beta} = - \mathbb{H}[C^{(xx)}_{\beta}].
\end{equation}

At this point, by comparing (\ref{eq:c_xy}) and (\ref{eq:c_yx}) we find another property
\begin{equation}
    C^{(yx)}_{\beta} = - C^{(xy)}_{\beta}.
\end{equation}

The covariance of the imaginary parts can be found following the same general logic.
First, we find by definition
\begin{multline}
\label{app-eq:Cyy-prefinal}
    C^{(yy)}_{\beta}
    = \mathbb{E} \left[ \sum_{\gamma = -\infty}^{\infty} \sum_{\delta = -\infty}^{\infty}
    h_{\gamma} h_{\delta} x_{\alpha-\gamma} x_{\alpha-\beta-\delta} \right] \\
    = \sum_{\gamma = -\infty}^{\infty} \sum_{\delta = -\infty}^{\infty}
    h_{\gamma} h_{\delta} C^{(xx)}_{\beta+\delta-\gamma}.
\end{multline}
It is easier to perform the remaining computations symbolically.
The important step is to recognize the Hilbert transforms and use the properties that were already used above in addition the property $\mathbb{H}[\mathbb{H}[f]] = -f$.
Then, (\ref{app-eq:Cyy-prefinal}) becomes
\begin{equation}
    \sum_{\delta = -\infty}^{\infty}
    h_{\delta} \mathbb{H} [C^{(xx)}_{\beta+\delta}]
    = - \mathbb{H} [\mathbb{H} [C^{(xx)}_{\beta}]]
\end{equation}
which gives the remaining covariance
\begin{equation}
    C^{(yy)}_{\beta} = C^{(xx)}_{\beta}.
\end{equation}

Thus, all the covariance properties of the analytic signal can be obtain from the known covariance of their real parts.

\section{Covariance matrix of analytic signal}
\label{app:cov-matrix-structure}
This Appendix gathers the computations that eventually lead to the dramatic simplification of the optimal detector from the form (\ref{eq:lnl}) to (\ref{eq:detector-prefinal-form}).
This simplification is possible due to the form of the covariance matrix of analytic signals.

To reveal the structure of the covariance matrix, let us first consider only a two-sample signal with the samples $z_1$ and $z_2$ sampled from a zero mean signal to simplify the computations without losing their generality.
This two-sample signal corresponds to $\mathbf{v} = (z_1 z_1^* z_2 z_2^*)^{\intercal}$ and the covariance matrix $V$ of the form
\begin{equation}
\label{eq:covmatrix}
V =
\mathbb{E}
\left[
\begin{array}{c c ;{2pt/2pt} c c}
z_1   z_1^* & z_1   z_1 &    z_1   z_2^* & z_1   z_2 \\
z_1^* z_1^* & z_1^* z_1 &    z_1^* z_2^* & z_1^* z_2 \\ \hdashline[2pt/2pt]
z_2   z_1^* & z_2   z_1 &    z_2   z_2^* & z_2   z_2 \\
z_2^* z_1^* & z_2^* z_1 &    z_2^* z_2^* & z_2^* z_2
\end{array}
\right].
\end{equation}
The lines highlight here the internal structure of the matrix.
It is easy to notice that the matrix has a block structure.
If we consider the matrix of more samples, the structure remains the same.
A single building block of such matrices reads
\begin{equation}
\label{eq:covmatrixblock}
\tilde{V}_{\alpha\beta} = \mathbb{E} \left[
\begin{array}{cc}
    z_{\alpha}   z_{\beta}^* & z_{\alpha}   z_{\beta} \\
    z_{\alpha}^* z_{\beta}^* & z_{\alpha}^* z_{\beta}
\end{array}
\right].
\end{equation}
In terms of these blocks, the general form of the covariance matrix has the following structure
\begin{equation}
  V =
\left[
\begin{array}{c c c c}
       & \vdots                          & \vdots                            & \\
\cdots & \tilde{V}_{{\alpha}\,{\beta}}   & \tilde{V}_{{\alpha}\,{\beta}+1}   & \cdots \\
\cdots & \tilde{V}_{{\alpha}+1\,{\beta}} & \tilde{V}_{{\alpha}+1\,{\beta}+1} & \cdots \\
       & \vdots                          & \vdots                            &
\end{array}
\right].
\end{equation}

After identifying the structure of the covariance matrix, the problem of studying its structure can be reduced to studying the structure of its blocks.
First, we can write down an equivalent from fo the matrix in (\ref{eq:covmatrixblock})
\begin{equation*}
\tilde{V}_{\alpha\beta} = \mathbb{E} \left[
\begin{array}{cc}
    z_{\alpha}  z_{\beta}^*  & z_{\alpha}   z_{\beta} \\
    (z_{\alpha} z_{\beta})^* & (z_{\alpha} z_{\beta}^*)^*
\end{array}
\right],
\end{equation*}
where the elements of the second row are the complex conjugated elements of the first row.
Thus, we should find the form only for the elements $\mathbb{E}[z_{\alpha} z_{\beta}^*]$ and $\mathbb{E}[z_{\alpha} z_{\beta}]$.

Let us first consider the element $\mathbb{E}[z_{\alpha} z_{\beta}^*]$.
By definition, this is the correlation function of the complex variables,
which for our case are the values of the analytic signals.
We use the usual notation $z_{\alpha} = x_{\alpha} + i y_{\alpha}$ for the complex number for further computations.
The explicit form of the element of the matrix can be written as the sum of the covariances between the signal components (see equation (\ref{eq:Rxx}))
\begin{multline}
\mathbb{E}[z_{\alpha} z_{\alpha+\gamma}^*] = \mathbb{E}[(x_{\alpha} + i y_{\alpha}) (x_{\alpha+\gamma} - i y_{\alpha+\gamma})] \\
= C^{(xx)}_{\gamma} - i C^{(xy)}_{\gamma} + iC^{(yx)}_{\gamma} + C^{(yy)}_{\gamma}.
\end{multline}

For the analytic signal, due to the connection of the real and imaginary part of the signals via the Hilbert transform, all the covariances are also connected to each other.
These connections are derived in the Appendix~\ref{app:covariances}.
If we use them, we obtain the general form of the element of the matrix
\begin{equation}
\label{eq:covmat-elements}
\mathbb{E}[z_{\alpha} z_{\alpha+\gamma}^*] = 2C^{(xx)}_{\gamma} + 2 i \mathbb{H}[C^{(xx)}_{\gamma}].
\end{equation}
The remaining covariance is known.
It is the one from the original real signal that is used to compute the analytic signal.

The same procedure gives zero for the second element of the matrix
\begin{equation}
\mathbb{E}[z_{\alpha} z_{\alpha+\gamma}] = 0.
\end{equation}

By collecting the results for the two elements, we find the general form of each block in the covariance matrix
\begin{equation}
\tilde{V}_{\alpha\,\alpha+\gamma} = \left[
\begin{array}{cc}
    C_{\alpha\,\alpha+\gamma} & 0 \\
    0  & C^*_{\alpha\,\alpha+\gamma}
\end{array}
\right].
\end{equation}
Here, for convenience, we introduced the shorthand notation
\begin{equation}
    C_{\alpha\,\alpha+\gamma} = 2C^{(xx)}_{\gamma} + 2 i \mathbb{H}[C^{(xx)}_{\gamma}].
\end{equation}

If we once again come to the simple matrix (\ref{eq:covmatrix}) for illustration, its explicit form becomes
\begin{equation}
\label{eq:covmatrixexpl}
V =
\left[
\begin{array}{c c ;{2pt/2pt} c c}
C_{11} & 0 &    C_{12} & 0 \\
0 & C^*_{11} &    0 & C^*_{12} \\ \hdashline[2pt/2pt]
C_{21} & 0 &    C_{22} & 0 \\
0 & C^*_{21} &    0 & C^*_{22}
\end{array}
\right].
\end{equation}

The matrix (\ref{eq:covmatrixexpl}) can be greatly simplified.
To see that, we should consider a complex vector similar to (\ref{eq:vvec}), but with a different arrangement of the elements
\begin{equation}
\label{eq:uvec}
  \mathbf{u} = (z_1 z_2 \dots z_N z_1^* \dots z_N^*)^{\intercal},
\end{equation}
The vectors $\mathbf{u}$ and $\mathbf{v}$ of the two different arrangements are connected via the permutation matrix
\begin{equation}
  \mathbf{u} = P \mathbf{v}.
\end{equation}
with the following explicit form
\begin{equation}
  P =
  \left[
  \begin{array}{c c  c c  c  c c  c c c}
    1 & 0 & 0 & 0 & \cdots \\
    0 & 0 & 1 & 0 & \cdots \\
    \vdots & \vdots & \vdots & \vdots & & & \vdots & \vdots & \vdots & \vdots \\
    &  &  &  &  & \cdots & 0 & 1 & 0 & 0 \\
    &  &  &  &  & \cdots & 0 & 0 & 1 & 0 \\ \hdashline[2pt/2pt]
    0 & 1 & 0 & 0 & \cdots \\
    0 & 0 & 0 & 1 & \cdots \\
    \vdots & \vdots & \vdots & \vdots & & & \vdots & \vdots & \vdots & \vdots \\
    & & & & & \cdots & 0 & 1 & 0 & 0 \\
    & & & & & \cdots & 0 & 0 & 0 & 1
  \end{array}
  \right].
\end{equation}
In each row and column of this matrix, only one element is one, and the remaining are zeros~\cite{bos1995}.

The bilinear form of type $\mathbf{v}^{\dag} V^{-1} \mathbf{v}$ that appeared in the log-likelihood ratio (\ref{eq:lnl}) can be transformed such that it uses the vectors $\mathbf{u}$ instead of $\mathbf{v}$~\cite{bos1995}.
To do this transformation, we use the property
\begin{equation}
    P^{-1}=P^{\intercal}
\end{equation}
of the permutation matrix~\cite{bos1995}.
Its usage provides the following
\begin{multline}
\label{eq:bilinear-form}
  \mathbf{v}^{\dag} V^{-1} \mathbf{v} = \mathbf{v}^{\dag} P^{\intercal} P V^{-1} P^{\intercal} P \mathbf{v} = \mathbf{u}^{\dag} P V^{-1} P^{\intercal} \mathbf{u} \\ =
  \mathbf{u}^{\dag} \left( P V^{-1} P^{\intercal} \right)^{-1} \mathbf{u} = \mathbf{u}^{\dag} U^{-1} \mathbf{u},
\end{multline}
where the matrix $U$ is defined as
\begin{equation}
  U = P V P^{\intercal}.
\end{equation}

If we find the form of the matrix $U$ associated to the matrix $V$ in (\ref{eq:covmatrixexpl}), we discover that the covariance matrix $U$ has a simpler block structure
\begin{equation}
\label{eq:umat-example-final}
U =
\left[
\begin{array}{c c ;{2pt/2pt} c c}
C_{11} & C_{12} &    0 & 0 \\
C_{21} & C_{22} &    0 & 0 \\ \hdashline[2pt/2pt]
0 & 0 &  C^*_{11} & C^*_{12} \\
0 & 0 &  C^*_{21} & C^*_{22}
\end{array}
\right],
\end{equation}
with two non-zero blocks on the main diagonal that are complex conjugate to each other, and two zero non-diagonal blocks.
For the larger matrices the structure remain the same.

Thus, after the transformation, the covariance matrix $U$ for an arbitrary signal has the following structure
\begin{equation}
  U =
  \begin{bmatrix}
    U_0 & 0\\
    0 & U_0^*
  \end{bmatrix}.
\end{equation}
From the illustrative example (\ref{eq:umat-example-final}), we can see that the matrix $U_0$ is comprised of the elements (\ref{eq:covmat-elements})
\begin{equation}
\label{eq:u0-structure}
    (U_0)_{\alpha,\beta} = 2 C^{(xx)}_{\beta-\alpha} + 2 i \mathbb{H}[C^{(xx)}_{\beta-\alpha}].
\end{equation}

The matrices $U$ and $U_0$ have the property $U^* = U^{\intercal}$ and $U_0^* = U_0^{\intercal}$ and both of them are hermitian matrices.
Also, due to the block structure of the matrix $U$, its inverse matrix has the simple form
\begin{equation}
\label{eq:Uinv}
  U^{-1} =
  \begin{bmatrix}
    U_0^{-1} & 0\\
    0 & (U_0^{-1})^*
  \end{bmatrix}.
\end{equation}
Since $U$ and $U_0$ are hermitian, this inverse matrix and its blocks $U_0^{-1}$ are also hermitian.

This new structure of the covariance matrix enables the simplification of the bilinear form (\ref{eq:bilinear-form}).
If we consider that form for the two vectors in the arrangement (\ref{eq:uvec})
\begin{equation}
    \mathbf{a} =
    \begin{pmatrix}
\mathbf{a_0}\phantom{^*} \\
\mathbf{a_0}^*
\end{pmatrix},\hspace{1cm}
\mathbf{b} =
    \begin{pmatrix}
\mathbf{b_0}\phantom{^*} \\
\mathbf{b_0}^*
\end{pmatrix}
\end{equation}
where $\mathbf{a}_0$ and $\mathbf{b}_0$ denote the corresponding sub-vectors, the bilinear form becomes
\begin{multline}
    \mathbf{a}^{\dag} U^{-1} \mathbf{b}
    = \mathbf{a}_0^{\dag} U_0^{-1} \mathbf{b}_0
    + \mathbf{a}^{\intercal} (U^{-1}_0)^* \mathbf{b}^* \\
    = \mathbf{a}_0^{\dag} U_0^{-1} \mathbf{b}_0
    + \left( \mathbf{a}_0^{\dag} U_0^{-1} \mathbf{b}_0 \right)^*.
\end{multline}
The zero non-diagonal blocks of the matrix $U$ eliminate half of the possible combinations of the sub-vectors.
This fact connected to the structure of the covariance matrix simplifies the optimal detector from (\ref{eq:lnl}) to (\ref{eq:detector-prefinal-form}).

\section{Karhunen-Lo\`{e}ve decomposition\\ for analytic signals}
\label{app:kl}
This appendix shows decomposition of the matrices $U_{H_0}$ and $U_{H_1}$ in (\ref{eq:detector-prefinal-form}) to the Karhunen-Lo\`{e}ve series.
Here we use the discrete form of the Karhunen-Lo\`{e}ve decomposition, that was published in~\cite{ray1970}.

Both matrices $U_{H_0}$ and $U_{H_1}$ in (\ref{eq:detector-prefinal-form}) have the same internal structure (\ref{eq:u0-structure}) and depend only on the covariance $C^{(xx)}_{\beta-\alpha}$ between the samples of the original real signal.

This covariance is different in the case of the two hypothesis.
If there is only noise, the covariance is simply the one given by (\ref{eq:noise-covariance})
\begin{equation}
    C^{(xx)}_{\beta-\alpha} = N_0 \delta_{\alpha\beta}.
\end{equation}
If in addition there is the signal, the covariance has its contribution and becomes
\begin{equation}
    C^{(xx)}_{\beta-\alpha} = K_{\alpha\beta} + N_0 \delta_{\alpha\beta},
\end{equation}
where $K_{\alpha\beta}$ is the covariance matrix (\ref{eq:initialcov}).
We can now write the explicit form of the elements of covariance matrices for the analytic signals
\begin{align}
\label{eq:uh0-matrix-equation}
    (U_{H_0})_{\alpha\beta} &= N_0 (\delta_{\alpha\beta} + i \mathbb{H}[\delta_{\alpha\beta}]),\\
    (U_{H_1})_{\alpha\beta} &= N_0 (\delta_{\alpha\beta} + i \mathbb{H}[\delta_{\alpha\beta}]) + (K_{\alpha\beta} + i \mathbb{H}[K_{\alpha\beta}]).
\end{align}

The Karhunen-Lo\`{e}ve decomposition is based on the solution of the matrix characteristic equation~\cite{ray1970}
\begin{equation}
    \sum_{i=\beta}^{N} U_{\alpha\beta} h_{\beta}
    = \eta^2 h_{\alpha},
\end{equation}
where $U_{\alpha\beta}$ is the covariance matrix ($(U_{H_0})_{\alpha\beta}$ or $(U_{H_1})_{\alpha\beta}$ in our case).
This equation can be rewritten in the vector form for further convenience
\begin{equation}
    U \mathbf{h} = \eta^2 \mathbf{h}.
\end{equation}
Now $\eta_i^2$ and $\mathbf{h}_i$ will be different eigenvalues and eigenvectors that are solutions of that equation.

Before providing the solution for that kind of equation, let us consider how the matrix $U_{H_0}$ acts on a vector of temporal samples of the analytic signal.
The Kronecker delta function of the elements means that the corresponding matrix is the unity matrix $\mathrm{I}$.
Using that fact and considering the discrete Hilbert transform as a corresponding matrix $\mathrm{H}$ (of finite size), we can transform (\ref{eq:uh0-matrix-equation}) to
\begin{equation}
    U_{H_0}
    = \frac{N_0}{2}(\mathrm{I} + i\mathrm{H}).
\end{equation}
If we check how the combination $(\mathrm{I} + i\mathrm{H})$ acts on the analytic signal $\mathbf{z} = \mathbf{x} + i\mathbf{y} = \mathbf{x} + i \mathrm{H}\mathbf{x}$, we observe
\begin{equation}
    (\mathrm{I} + i\mathrm{H}) (\mathbf{x} + i \mathrm{H}\mathbf{x})
    = 2(\mathbf{x} + i \mathbf{y}),
\end{equation}
where we used the property $\mathrm{H}[\mathrm{H}[f]] = -f$.
Thus, this combination effectively acts similarly to the unitary matrix.
It only doubles the scale of the original vector.
That means that the solution for the equation
\begin{equation}
    U_{H_0} \mathbf{h} = \eta^2 \mathbf{h}
\end{equation}
can be any vector $\mathbf{h}$.
The eigenvalues for this equation are equal to the constant in (\ref{eq:uh0-matrix-equation}) multiplied by a factor of two, as shown above,
\begin{equation}
    \eta^2_{H_0} = 2 N_0.
\end{equation}

To solve the characteristic equation for the matrix $U_{H_1}$ one should know the particular form of the covariance matrix, $K_{\alpha\beta}$, which depends on the particular problem in question.
However, the general solution can be written since we know how the combination $(\mathrm{I} + i\mathrm{H})$ acts on an arbitrary vector.
If the eigenvalues $\eta_i^2$ and eigenvectors $\boldsymbol{\phi}_i$ are the solution of the characteristic equation in the vector form
\begin{equation}
    (K + i \mathbb{H}[K])\boldsymbol{\phi}_i = \eta_i^2\boldsymbol{\phi}_i,
\end{equation}
then the Karhunen-Lo\`{e}ve decomposition of the matrix $U_{H_1}$ is
\begin{equation}
    U_{H_1} = \sum_{i=1}^{N} \left(2 N_0 + \eta_i^2\right) \boldsymbol{\phi}_i \boldsymbol{\phi}_i^{\dag}.
\end{equation}
Since the matrix $U_{H_0}$ is such that any vectors can be used for the decomposition, we use the same vectors as for $U_{H_1}$
\begin{equation}
    U_{H_0} = \sum_{i=1}^{N} 2 N_0 \boldsymbol{\phi}_i \boldsymbol{\phi}_i^{\dag}.
\end{equation}

By using the fact that the vectors $\boldsymbol{\phi}_i$ are orthogonal by definition of the transform, the result for the inverse matrices (\ref{eq:KL-decomposition}) follows straightforwardly.

\section{Simplification of the internal product pairs}
\label{app:simplification}
For completeness, we derive in this appendix the simplifications of the terms (\ref{eq:lr-initial}) and (\ref{eq:ld-initial}).

First, we can right away transform (\ref{eq:lr-initial}) to
\begin{equation}
\label{app-eq:lr}
    l_{\mathrm{R}} = \frac{1}{2 N_0} \sum_{i=1}^{N} \frac{\eta_i^2}{2 N_0 + \eta_i^2}
    \mathbf{r}^{\dag} \boldsymbol{\phi}_i \boldsymbol{\phi}_i^{\dag} \mathbf{r}
\end{equation}
by summing up all the terms.

Then, we can use the fact that the combinations of vectors
$\mathbf{r}^{\dag} \boldsymbol{\phi}_i \boldsymbol{\phi}_i^{\dag} \mathbf{r}$,
$\mathbf{m}^{\dag} \boldsymbol{\phi}_i \boldsymbol{\phi}_i^{\dag} \mathbf{r}$,
$\mathbf{r}^{\dag} \boldsymbol{\phi}_i \boldsymbol{\phi}_i^{\dag} \mathbf{m}$
in (\ref{eq:lr-initial}) and (\ref{eq:ld-initial}) are pairs of internal products.
By transforming them, we can further simplify the form of the detector.

First, we can transform $\mathbf{r}^{\dag} \boldsymbol{\phi}_i \boldsymbol{\phi}_i^{\dag} \mathbf{r}$ by exploiting the fact that the internal product does not change by the operation of transposition
\begin{equation}
    (\mathbf{r}^{\dag} \boldsymbol{\phi}_i )( \boldsymbol{\phi}_i^{\dag} \mathbf{r})
    = |\boldsymbol{\phi}_i^{\dag} \mathbf{r} |^2.
\end{equation}
This further simplifies (\ref{app-eq:lr}) and we obtain (\ref{eq:lr-after-simplification}).

If we consider (\ref{eq:ld-initial}), we see that the pair of products is simply
\begin{equation}
    (\mathbf{r}^{\dag} \boldsymbol{\phi}_i)(\boldsymbol{\phi}_i^{\dag} \mathbf{m})
    = (\boldsymbol{\phi}_i^{\dag} \mathbf{r})^{\dag}(\mathbf{m}^{\dag} \boldsymbol{\phi}_i)^{\dag}.
\end{equation}
The complex conjugate part of (\ref{eq:lnl-prefinal}) for the same product of the same term is
\begin{equation}
    (\boldsymbol{\phi}_i^{\dag} \mathbf{r})^{\intercal}(\mathbf{m}^{\dag} \boldsymbol{\phi}_i)^{\intercal},
\end{equation}
that is equal to
\begin{equation}
    (\boldsymbol{\phi}_i^{\dag} \mathbf{r})(\mathbf{m}^{\dag} \boldsymbol{\phi}_i),
\end{equation}
since transposition does not change a number.
This is the same pair of the internal products as in the first term in (\ref{eq:ld-initial}).
Thus, we can take the complex conjugate term corresponding to the second term in (\ref{eq:ld-initial}) and transform its internal products to $(\boldsymbol{\phi}_i^{\dag} \mathbf{r})(\mathbf{m}^{\dag} \boldsymbol{\phi}_i)$ and use it in the equation for $l_{\mathrm{D}}$ and term itself place in the complex conjugate part of (\ref{eq:lnl-prefinal}).
After performing all these actions, we arrive at (\ref{eq:ld-after-simplification}).
}

\bibliographystyle{ieeetr}
\bibliography{references}

\begin{IEEEbiographynophoto}{Vladimir Lenok}
was born in Angarsk, Russia, in 1990.
He received the Diploma of Engineer-Physicist degree from NRNU MEPhI, Moscow, Russia, in 2013
and the Dr. rer. nat. degree from the Karlsruhe Institute of Technology, Karlsruhe, Germany, in 2021.

He joined the Bielefeld University in 2022 and he is currently a Postdoctoral Researcher.

Dr.\,Lenok is a member of the German Astronomical Society, the European Astronomical Society, and Junior Member of the International Astronomical Union.

\end{IEEEbiographynophoto}

\end{document}